\documentstyle[11pt]{article}
\begin{document}
\newcommand{\ECM}{\em Departament d'Estructura i Constituents de la
Mat\`eria
                  \\ Facultat de F\'\i sica, Universitat de Barcelona \\
                     Diagonal 647, 08028 Barcelona, Spain}

\def\thefootnote{\fnsymbol{footnote}}
\pagestyle{empty}
{\hfill \parbox{6cm}{\begin{center} UB-ECM-PF 95/14\\
                                    hep-ph/9507258 \\
                                    July 1995
                     \end{center}}}
\vspace{1.5cm}

\begin{center}
\large{Constraints on Chiral Perturbation Theory Parameters
 from QCD Inequalities}

\vskip .6truein
\centerline {Jordi COMELLAS\footnote{e-mail: comellas@sophia.ecm.ub.es},
 Jos\'e Ignacio LATORRE\footnote{e-mail: latorre@sophia.ecm.ub.es}
and Josep TARON\footnote{e-mail: taron@ifae1.ecm.ub.es}}
\end{center}
\vspace{.3cm}
\begin{center}
\ECM
\end{center}
\vspace{1.5cm}

\centerline{\bf Abstract}
\medskip
We explore some of the
 constraints imposed by positivity of the QCD measure (Weingarten's
inequalities) on the parameters defining chiral perturbation theory.
 We find,
in particular, that $2 m_q (\mu)\leq B_0(\mu) Z(\mu)$.
The use of further properties of the
exact fermion propagator yields information on
some higher order parameters.

\newpage
\pagestyle{plain}

\section{QCD inequalities}
 \label{QCD}
 The long-distance realization of QCD is presently assumed to be
described by chiral perturbation theory \cite{GL}.
 This idea is solidly based on the way QCD symmetries are implemented
at low energies.
 In this letter we shall explore  first principle
constraints on  QCD through some
inequalities inherent to vector-like gauge theories.

To set up our approach we first briefly review Weingarten's original idea
\cite{WE}.
Let us
consider an Euclidean
 vector current, $V^a_\mu(x)={\rm i} \bar\psi(x) \gamma_\mu
{\lambda^a\over 2}\psi(x)$, where $\lambda^a$ carries $SU(n_f)$ flavor indices.
 The Euclidean two-current correlator is
\begin{equation}
 \langle V^a_\mu(x)\ V^b_\nu(0)\rangle = \int d\mu\ {\rm Tr}
\left(S_{x,0}\gamma_\mu {\lambda^a\over 2}S_{0,x}\gamma_\nu
{\lambda^b\over 2}\right),
\end{equation}
where $S_{x,0}$ corresponds to the exact fermionic propagator
in the presence of a classical gluonic background and
 $d\mu$ stands for the gluonic measure, including the fermionic
determinant, which we now assume positive and postpone its discussion.
Weingarten made the observation that the Cauchy-Schwarz matrix inequality,
$\left|{\rm Tr} (UV^\dagger)\right|^2\leq
 {\rm Tr} (UU^\dagger){\rm Tr} (VV^\dagger)$,
is applicable to the spinor trace. Thus, given a positive measure,
the following chain of
reasoning holds :
\begin{equation}
\left|\langle V^a_\mu(x) \ V^a_\nu(0)\rangle \right| \leq
\int d\mu\ \left| {\rm Tr}\left(S_{x,0}\gamma_\mu {\lambda^a\over 2}
S_{0,x}\gamma_\nu{\lambda^a\over 2}\right)\right|
\leq
\int d\mu\ {\rm  Tr} \left( S_{x,0} \gamma_5 {\lambda^a\over 2}S_{0,x}
\gamma_5 {\lambda^a\over 2}\right) \ ,
\end{equation}
where we have used the Euclidean properties $\gamma_\mu^\dagger=\gamma_\mu$
and
$S_{x,0}^\dagger=\gamma_5 S_{0,x}\gamma_5$. Note that no summation on
$a$ is implied and that the inequality holds no matter what space
indices are taken. The last expression
is manifestly positive as it is the square
of the absolute value of a complex matrix. Noticing that this
corresponds to the  correlator of two bare pseudoscalar currents
 ($P_0^a(x) = {\rm i} \bar \psi \gamma_5 {\lambda^a\over 2}\psi$),
 it follows that
\begin{equation}
\label{vvppone}
\left| \langle V^a_\mu(x)\ V^a_\nu(0)\rangle \right| \leq
\langle P_0^a(x) P_0^a(0)\rangle = Z(\mu) \langle P^a(x) P^a(0)\rangle \ ,
\end{equation}
where $P_0^a=Z^{1\over 2}(\mu) P^a(\mu)$, and $Z(\mu)$ is the quark
renormalization constant introduced as $m_0=Z^{-{1\over 2}}(\mu) m(\mu)$.
This composite operator renormalization of the pseudo-scalar current is
dictated by the Ward Identity that relates the non-renormalization of the
derivative of the axial current to the product $m P^a$.
Weingarten showed  that this result, when
combined with a narrow resonance approximation, leads to

\begin{equation}
a(x) e^{-m_\rho |x|} \leq b(x) e^{-m_\pi |x|}\ ,
\end{equation}
where $a(x)$ and $b(x)$ are polynomials in $x$. Thus,
an
elegant inequality between masses follows
\begin{equation}
m_\rho\geq m_\pi ,
\end{equation}
since no power law can beat  the exponential decay as $|x|\to\infty$.

The very same inequality may be applied to QED, yielding the result that
orthopositronium is heavier than parapositronium. Furthermore, whatever
quantum numbers are used at the outset, the inequality is always
 bounded by the
pseudo-scalar correlator, if just the correlator can be written as a
single trace of two propagators closing a fermion loop.
 This shows that the pion is the lowest lying
meson of the QCD spectrum.

Let us now come back to the discussion of the  positivity of the
measure. Formally, the gluonic measure which includes the integration
over fermionic variables is
\begin{equation}
d\mu \equiv \left[d A^a_\mu(x)\right] e^{-{1\over 4} \int
F^a_{\mu\nu}F^a_{\mu\nu} } \left({\rm det}(D\!\!\!\! /+m)\right)^{n_f}\ ,
\end{equation}
where $n_f$ is the number of flavors.
This expression needs regularization. Weingarten \cite{WE}
 argued that the lattice
provides a regularization which corresponds to a product of Haar
positive measures
on each link. Positivity would then hold uniformly in the continuum,
 infinite volume  and zero quark mass limits.

Later, Vafa and Witten \cite{VW}
 used the fact that non-zero eigenvalues of the
gauged Dirac operator are paired  via multiplication by $\gamma_5$
\begin{equation}
  {\rm i} D\!\!\!\! / \ \phi =\lambda \phi\quad\longrightarrow
\quad
{\rm i} D\!\!\!\! / \ \gamma_5 \phi= -\lambda\gamma_5 \phi\ .
\end{equation}
As a consequence,
\begin{equation}
{\rm det}(  D\!\!\!\! / \ +m)=\prod_{\lambda> 0}
(\lambda^2 +m^2) \prod_{\lambda=0} m\ ,
\end{equation}
which is formally positive definite.
A non-perturbative gauge-invariant
parity-preserving regularization is likely to be supplied by
proposals which involve higher-derivatives, as in ref. \cite{AF}.

Most remarkably, no violation of any of the inequalities derived from
positivity of the measure is known. This includes a large number of different
applications analyzed in the literature.

A second set of basic  inequalities was put forward by Witten \cite{WI}
using
the property that $E$, the piece of the propagator that commutes with
$\gamma_5$,
\begin{equation}
\label{operatore}
E\equiv S+\gamma_5\ S\ \gamma_5={2 m\over -D\!\!\!\! / ^{\, 2}+m^2}\quad
,\quad S={1\over D\!\!\!\! / + m}\quad ,
\end{equation}
is a positive operator. Witten used this fact to explain the positivity
of $m_\pi^+{}^2-m_\pi^0{}^2$.

We shall try to exploit both kind of inequalities to put constraints on
long-distance realizations of QCD.

\section{Weingarten's inequalities in chiral perturbation theory}
\label{wcpt}

As a first exercise we consider the inequality given in Eq. (\ref{vvppone})
in the framework of $SU(2)$ chiral perturbation theory, with $m_q=m_u=m_d$.
We need to recall that to lowest order
\begin{equation}
V^a_\mu(x)= {\rm i} f^{abc} \pi^b(x) \partial_\mu \pi^c(x)+\dots \quad,\quad
P^a(x)= -{\rm i} B_0 f_\pi \pi^a(x)+\dots
\end{equation}
 An immediate coordinate
space computation for the Euclidean
 vector-vector correlator gives to first order (no sum over $a$)
\begin{equation}
\langle V^a_\mu(x) \ V^a_\nu(0)\rangle   =
 {1\over 8 \pi^4} \left({-4 x_\mu x_\nu+2
\delta_{\mu\nu}x^2\over x^6}
m_\pi^2 K_1^2
 + {\delta_{\mu\nu}\over x^3} m_\pi^3
K_1 K_0+{x_\mu x_\nu\over x^4} m_\pi^4 \left(
K_0^2-K_1^2\right) + \dots\right)
\end{equation}
and, for the pseudo-scalar-pseudo-scalar one,
\begin{equation}
\langle P^a(x) P^a(0)\rangle= {1\over 4 \pi^2}f_\pi^2 B_0^2 {m_\pi\over
x} K_1+ \dots\
\end{equation}
where   $K_n\equiv K_n(m_\pi|x|)$ are Bessel functions and  the dots stand for
contact terms and higher order contributions.
We are now          free  to check Weingarten's inequality
 choosing at our  convenience any particular direction
of $x_\mu$ or, if preferred,  summing over $\mu=1,..,4$. In all cases, we
obtain
\begin{equation}
\alpha(x) e^{-2 m_\pi |x|} \leq \beta(x) e^{-m_\pi |x|},
\end{equation}
where $\alpha(x)$ and $\beta(x)$ are polynomials in $x$. The inequality
holds at $x\to \infty$ due to the correct description of the pion
content of each current. The exponential decay associated
to the two-pion threshold is bounded by the one-pion exchange observed
in the pseudo-scalar channel. It is also arguable that the r.h.s. of
the inequality is order $f_\pi^2$ whereas the l.h.s. is order 1, thus
sub-leading.

The massless limit of the above analysis yields
\begin{equation}
{1\over 4\pi^4} {1\over x^6}\leq {1\over 4\pi^2} f_\pi^2 B_0(\mu)^2
Z(\mu){1\over x^2}
\end{equation}
which is correctly obeyed as $x\to\infty$ due to the faster decay
of the product of two propagators as compared to a single one.
Nevertheless, neither the massive nor the massless inequalities are
valid for any $x$. Both are violated at a distance of the order of
the inverse of the pion mass, which is a hint at the need of higher order
corrections when $x$ is decreased.

A word is needed about subtractions. The process of renormalization
can be synthesized saying that bare amplitudes are transformed into
distributions by correcting just their singular points.
 In our case, the pseudo-scalar channel produces right away
a  {\sl bona fide} distribution whereas  the vector one does not. The
product of two propagators is not a distribution, due to the $x=0$
singularity.
Any regularization takes care of this problem by
subtracting  a contact term. Subtractions are thus of the form, {\sl e.g.}
in dimensional regularization,
\begin{equation}
{1\over \epsilon} \Box \delta^4(x)\quad, {m_\pi^2\over \epsilon}
\delta^4(x)
\
, ...
\end{equation}
Therefore,
our discussion on the long-distance exponential decay of two-point correlators
is clean and free of subtraction ambiguities. We are discussing the
physics of the non-local part of the amplitude which remains unchanged
along the renormalization process at this order.

For the sake of completeness, let us note that the analog of a standard
momentum space renormalized amplitude\footnote{
Note that $\log p^2$ is to be understood as $p^2 {\log p^2\over p^2}$
in the
sense of distributions, the first $p^2$
being replace by $-\Box$ which acts by parts when necessary.}
\begin{equation}
\lambda+\lambda^2 \log {p^2\over \nu^2}+\dots
\end{equation}
takes the following form in coordinate space \cite{FJL}
\begin{equation}
\lambda \delta^4(x) +\lambda^2 \Box{\log x^2\nu^2\over x^2}+\dots
\end{equation}
Changes of renormalization scheme,
$\nu\to \nu'$, are absorbed by redefinitions of the
coupling constant.
A massless amplitude  issuing from a non-renormalizable perturbation
theory will read an expression of the kind
\begin{equation}
 \delta^4(x)+{1\over f_\pi^2}
 \Box\delta^4(x)+\Box{\log x^2\nu^2\over x^2}
+{1\over f_\pi^2}\Box\Box{\log x^2\nu^2\over x^2}+\dots
\end{equation}
Only away from contact, boxes are allowed to act on the logs and the
expansion parameter takes the form of ${1\over x^2 f_\pi^2}$.
In chiral perturbation theory, contact terms associated with $L$'s do
 appear. Again, those are cleanly decoupled of our discussion above.

We now return to our exploitation of Weingarten's inequalities. As seen
in our first example, the long-distance two-pion decay gives the clue
to understand the fulfillment of the $VV<PP$ inequality. It is clear that
a more constraining result can only emerge from an inequality involving
two channels which are mediated by only one pion. This is the case of
\begin{equation}
\label{goodine}
\left|\langle A^a_\mu(x)A^a_\nu(0) \rangle\right|
 \leq Z(\mu) \langle P^a(x) P^a(0)\rangle\ ,
\end{equation}
(a narrow resonance approximation would tell us that the $A_1$
 meson is
heavier than the pion). In chiral perturbation theory  we obtain
($A^a_\mu(x)=f_\pi \partial_\mu \pi^a(x)$)
\begin{equation}
\label{fullmb}
\left| f_\pi^2 \partial_\mu\partial_\nu \left(
{m_\pi\over x}K_1(m_\pi x)\right)\right|\leq f_\pi^2 B_0(\mu)^2
Z(\mu){m_\pi\over x}
K_1(m_\pi x)\ .
\end{equation}
Taking $\mu=\nu=1$ and $x_2=x_3=x_4=0$, $x_1=y$,
we observe that
both channels decay at the same exponential and leading power rate.
 The constraint we
obtain is, thus, an inequality between
\begin{equation}
\label{mb}
m_\pi^2\leq B_0^2(\mu)Z(\mu)\ .
\end{equation}

Before entering the philosophical discussion of the result, let us
note that the inequality (\ref{fullmb}) is again violated when $y$  approaches
$1/m_\pi$, a consistent sign of the need for higher order corrections
at shorter distances.
The massless limit is also verified
since the axial channel is then reduced to a contact term. At variance
with the $VV<PP$ case, both sides of the inequality are of the same
order in $f_\pi$.

How should Eq. (\ref{mb}), and alike, be interpreted?

If the chiral expansion reproduces, order by order, a good
approximation to the QCD correlators at large distances, the
inequalities become constraints among the parameters in the effective
theory, which, in principle, are calculable in QCD ({\sl e.g.} on the
lattice). Observables  such as $m_\pi^2$ are expressed as functions of
these parameters. In general, the inequalities we found are not among
observables
\footnote{
We thank A. Manohar for an observation which triggered this discussion.}.

In Eq. (\ref{mb}), $m_\pi^2$ is not a parameter of the chiral lagrangian.
If one assumes that the order parameter $B_0$ gives the main contribution
to $m_\pi^2=2 m_q B_0+ O(m_q^2)$, it sets $m_q=O(m_\pi^2)$, hence second
order in chiral power counting. Therefore, Eq. (\ref{mb}) reads
\begin{equation}
\label{mb2}
2 m_q(\mu)\leq B_0(\mu)Z(\mu)\ .
\end{equation}
This is a constraint of QCD on the relative strength of the explicit
breaking of chiral symmetry, driven by the quark mass $m_q$, versus
the spontaneous breaking. It is no longer an assumption as it is imposed by
the vector-like structure of the theory. Notice also that both sides of the
inequality run similarly under the change of subtraction point.
Thus, by multiplying the inequality by $Z^{-{1\over 2}}(\mu)$ we obtain
a constraint between two renormalization group invariant quantities, often
written as $2 \hat m\leq \hat B$.

Other scenarios are possible in which a different chiral-counting for
$m_q$ is required. In the framework of Generalized Chiral Perturbation
Theory\cite{FSS}\cite{KS}
 one could have assumed that the main contribution to $m_\pi^2$
comes from terms up to quadratic order in $m_q$. In the limit of isospin
symmetry
\begin{equation}
m_\pi^2=2 m_q B_0+4 m_q^2 (A_0+2 Z_0^s)+O(m_q^3)
\end{equation}
(see refs. \cite{FSS}\cite{KS}
 for definitions of $A_0$ and $Z_0^s$) and, here,  $m_q$ does not
count as $m_\pi^2$ anymore.
 If $B_0$ were small enough, then $m_q=O(m_\pi)$, which
amounts to a re-shuffling of chiral orders in the standard expansion in
the explicit breaking sector. The axial-axial correlator is also
modified and Eq. (\ref{mb2}) reads
\begin{equation}
\label{gmb2}
Z(\mu) \left| {B_0\over 2  m_q} + A_0+2 Z_0^s\right| \geq 1
\end{equation}
The constraint is nontrivial. If $B_0$ is small, it still relates the
allowed values of $A_0$ and $Z_0^s$ in a way independent of the value of $m_q$
 (in a mass independent subtraction scheme).

We wish to emphasize that the relative importance of the order
parameters $B_0$, $A_0$, $Z_0^s$ and possibly others has a unique answer
in QCD, and it should be properly incorporated in order to have a
meaningful expansion. It is QCD what sets the chiral power counting of
$m_q$, which, if not properly taken into account, may result in a
violation of the inequality, order by order in the expansion. A
re-summation to all orders is then required to verify the inequality again.

Similarly, if $m_q$ is very big, {\sl e.g.} as the charm quark mass, a
chiral symmetric theory incorporating the charmed flavor is not expected
to be a good approximation. The violation of Eq. (\ref{mb2}) is apparent.
 However,
in a situation where $m_q$ is reasonably small so that
a chiral expansion is expected to work, if $B_0$
turns out to be even smaller, a violation of
  Eq. (\ref{mb2}) is a sign from QCD that the usual expansion is not the
appropriate one. Yet, on physical grounds, a generalized expansion is
expected to exist. Then Eq. (\ref {gmb2}) may not be violated and
should be regarded as a rigorous constraint.

We consider now the renormalization of our inequality to one loop.
At this point, we have to understand that the mass of the pion we
have been using so far is bare, $m_\pi=m_0$, and that the inequality we
were discussing is $m_0^2\leq B_0^2$. The one-loop renormalization
turns out to be particularly simple  as all one-loop graphs come from
three different sources: i) tadpoles associated to the composite
operator structure of the currents, ii)           tadpoles coming
through the expansion of the ${\cal L}_2$ term
in the chiral lagrangian to next order and
iii) insertions of  $L$'s, coming
form ${\cal L}_4$. None of these renormalizations
change the spatial behavior
of the correlators but only its parameters. All contributions are
finite due to the non-renormalization associated to the partial
conservation of the axial current. When finite parts are gathered
we get
\begin{equation}
m_\pi^2 \leq Z(\mu)B_0(\mu)^2 Z_m
\end{equation}
where (for SU(2))
\begin{equation}
Z_m={m_\pi^2\over m_0^2}= 1-{8 m_\pi^2\over f_\pi^2} \left(
2 L_4^r+L_5^r-4 L_6^r-2 L_8^r+{m_\pi^2\over 32 \pi^2 f_\pi^2}
\log {m_\pi^2\over \mu^2}\right)
\end{equation}
and, therefore, the bare inequality remains unaltered. Note that $Z_m$ is
not $\mu$ dependent.

At two-loop order, we do encounter a change of the behavior of the
correlators leading to the first appearance of three-pion thresholds.
All other diagrams are combinations of tadpoles that again will
only renormalize the parameters of the inequality.

Nothing prevents us to play at will with non-diagonal correlators in
Weingarten's setting. We have done so for $\langle A P\rangle
\leq \langle P P\rangle$ and obtained identical results to the above
ones. Full consistency of Ward Identities demanded so.
We have also performed a number of  checks involving three-point amplitudes.
Using a combination of Cauchy-Schwarz and H\"older inequalities we
have proven {\sl e.g.}
\begin{equation}
\left| \langle V^a_{\mu,x} P^b_y P^c_z\rangle\right|\ \leq\ Z(\mu)^{1\over 2}
\sqrt{2}\ \left|\varepsilon^{abc}\right| \
\left(\langle P^a_zP^a_x\rangle
\langle P^a_zP^a_y\rangle\langle P^a_yP^a_x\rangle\right)^{1\over 2}\ ,
\end{equation}
where we have used a shorter but obvious notation.
This inequality demands that the amplitude must decay exponentially
at least as
half of the mass of the pion in each pair of points. Although
non-trivially,
this is automatically fulfilled in chiral perturbation theory as we
have checked.   Collinear configuration of the three space points do
saturate the inequality between  exponential decays but the power
laws manage to keep the result safe.

\section{Weingarten's inequalities in momentum space}

The inequalities we have found among chiral lagrangian parameters are
independent of the technique used to derived them.
We here digress momentarily to reobtain our results in momentum space
so as to make an easiest contact with the way results are often
presented in the literature.

The proof of Weingarten's inequalities, as sketched in the introduction,
does not go through in momentum space for a good reason: unlike in
coordinate space, the two fermion propagators $S_k$ and $S_{q+k}$
differ in their arguments by a momentum insertion $q$, and are thus
unrelated. Naive manipulation of Cauchy-Schwarz inequality will not
reconstruct the  traces conveniently.

There exists, however, an inequality in momentum space at any value of
Euclidean $Q^2$ for each inequality in coordinate space. The proof starts
 from the
observation  that for a given  operator $B$
\begin{equation}
\label{spectral}
\int d^3 x \langle 0|B(-{\rm i} \tau,\vec x) B(0,\vec 0)|0\rangle =
\int_0^\infty dE e^{-E\tau} \rho(E^2)\ ,
\end{equation}
which holds for any value of Euclidean time $\tau>0$ and where $\rho$ is
the spectral function defined (in Minkowski space) as
\begin{equation}
\rho(q^2) =\sum_\Gamma (2 \pi)^3 \delta^4(q-p_\Gamma)
\langle 0|B(0,\vec 0)|\Gamma\rangle \langle\Gamma | B(0,\vec 0)|0\rangle\ ,
\end{equation}
where the sum is extended to all possible intermediate states with
suitable quantum numbers.

Since the integral in $\vec x$ has positive measure, Weingarten's
inequalities in Euclidean coordinate  space translate into inequalities
between transforms of spectral functions if the spatial integral
exists. The generic Weingarten's inequality for $B$ bilinear in
quark fields $
\left| \langle B(x) B(0)\rangle\right| \leq \langle P(x) P(0)\rangle
$
becomes
\begin{equation}
\label{phiine}
\left| \Phi(\tau)\right| \leq  \Phi_P(\tau)\ ,
\end{equation}
where $\Phi(\tau)$ is the Laplace transform of the spectral function
$\rho(E^2)$, as in Eq. (\ref{spectral}).

The momentum (Minkowski) space correlator
\begin{equation}
\Pi(q^2)={\rm i} \int d^4x\ e^{{\rm i} q x} \langle 0| T B(x) B(0)| 0\rangle\ ,
\end{equation}
verifies a dispersion relation which may need subtractions. For instance,
a twice subtracted dispersion relation is of the form ($q^2=-Q^2<0$)
\begin{equation}
\label{drexample}
\Pi(Q^2)=\Pi(0)+\Pi'(0) Q^2 +(Q^2)^2 \int_0^\infty dE^2\
{\rho(E^2)\over (E^2)^2 (E^2+Q^2)}\ .
\end{equation}
Our aim is to establish inequalities among the functions $\Pi(Q^2)$
at Euclidean momenta. The strategy is to convolute
inequality (\ref{phiine})
 with positive functions $F(\tau)>0$. It turns out that the
functions $F_0=1+\cos Q\tau$, $F_1=1-\cos Q\tau$,
$F_2=({Q\tau\over 2})^2-\sin^2({Q\tau\over 2})$ and
$F_3= {1\over 3} ({Q\tau\over 2})^4- ({Q\tau\over 2})^2+\sin^2({Q\tau\over 2})
$ are positive and lead to (we omit $Z(\mu) $ factors)
\begin{eqnarray}
\label{momine}
&\left| \Pi(Q^2)+\Pi(0)\right| \leq \Pi_P(0)+\Pi_P(Q^2)\ ,\cr
&\left| \Pi(0)-\Pi(Q^2)\right| \leq \Pi_P(0)-\Pi_P(Q^2)\ ,\cr
&\left| \Pi(Q^2)-\Pi(0)-Q^2 \Pi'(0)\right| \leq \Pi_P(Q^2)-\Pi_P(0)
-Q^2 \Pi_P'(0)\ ,\cr
&\left| {1\over 2} (Q^2)^2 \Pi''(0) +Q^2 \Pi'(0)+\Pi(0)- \Pi(Q^2)\right|
 \leq {1\over 2} (Q^2)^2 \Pi_P''(0) +Q^2 \Pi_P'(0)+\Pi_P(0)- \Pi_P(Q^2)
\ ,\cr
\end{eqnarray}
which are inequalities that apply when zero, one, two and three
subtractions are required. The generalization to any number of
subtractions is immediate. In the cases where $\langle B(x) B(0)\rangle$
is positive for all $x$, the inequalities hold without the need for
absolute values in the l.h.s. of (\ref{momine}).

Notice that these inequalities involved subtracted correlators.
Mathematically, this is due to the fact that the inverse Laplace
transform of ${1\over Q^2+E^2}$ is $2 \cos Q\tau$, which is not a
positive function. Remarkably enough, there exist positive functions
($F_0,...,F_n$) that enable the extraction of the desired inequalities
by taking proper care of the subtractions needed to end up with a
convergent integral of $\rho(E^2)$.

Convolutions with even powers of $\tau$ also furnish inequalities among
derivatives of $\Pi(Q^2)$ at $Q^2=0$. Again, care must be taken of the
subtractions: the more subtractions needed, the higher the derivative of
$\Pi$ is to be considered.

Let us consider the example of the vector correlator. In Minkowski space
one has, in the limit of exact isospin symmetry,
\begin{equation}
{\rm i} \int d^4x\ e^{{\rm i} q x}\langle0| T V^a_\mu(x) V^a_\nu(0)|0\rangle =
\left( q_\mu q_\nu-q^2 g_{\mu\nu}\right) \Pi_V(q^2)
\ .
\end{equation}
We proceed by defining the standard spectral function
\begin{equation}
\left( q_\mu q_\nu-q^2 g_{\mu\nu}\right) \rho_V(q^2)
=
\sum_{\Gamma} (2\pi)^3 \delta^4(q-p_\Gamma)
 \langle 0| V_\mu(0,\vec 0)|\Gamma\rangle
 \langle \Gamma| V_\nu(0,\vec 0)|0\rangle\ .
\end{equation}
Similar dispersion relations to that in Eq. (\ref{drexample}) relate
$\Pi_V$ to $\rho_V$. In QCD, $\Pi_V$ requires one
subtraction ( see for instance \cite{FNR}).
Working out all inequalities is now straightforward,
\begin{equation}
\label{mompre}
 Q^2 \left| \Pi_V'{}^{(1)}(Q^2) -\Pi_V'{}^{(1)}(0)\right| \leq
\Pi_P(Q^2)-\Pi_P(0)-Q^2 \Pi_P'(0)\ .
\end{equation}
Moreover, the application of
an even-power convolution is even simpler, yielding the result
\begin{equation}
\label{momres}
 \left| \Pi_V'(0)\right| \leq {1\over 2} \Pi_P''(0)\ ,
\end{equation}
for $F(\tau)=\tau^4$.
 Notice that this procedure removes contact terms related to
subtractions. This reminds us that in
coordinate space the same was automatically done by just analyzing
correlators at non-zero distances.

Using a similar result to Eq. (\ref{momres}) for the non-transverse
part of the axial correlator, the inequality
 $m_\pi^2\leq B_0(\mu)^2Z(\mu)$ emerges again.

Once chiral perturbation theory is set
up, the inequalities we get for the parameters of the lagrangian
cannot depend on which space the inequalities are treated. It is a
technicality to  pass from one space to the other, the physical
content remaining the same. Coordinate space allows a far simpler
analysis due to the trivial decoupling of subtractions at large distances.

\section{Witten's inequalities in chiral perturbation theory}

As an initial remark we note that the positivity of the operator
$E$ is formally
related to the sign of  the condensate, $\langle \bar q
q\rangle = -{1\over V}\int d\mu\ {\rm Tr} E$, where $V$ is the volume
of space. This observation is further related
to the infrared limit of the spectral density of the quark propagator
(see ref. \cite{BCS}).

Let us now concentrate again on two-point correlators. We follow
 Witten's argument \cite{WI} and define
 the matrix $E_{x,0}\equiv S_{x,0}+\gamma_5 S_{x,0}\gamma_5$
which commutes with $\gamma_5$ and corresponds to the matrix
element of the positive operator $E$ in Eq. (\ref{operatore}).
Then,
\begin{equation}
\int d\mu\ {\rm Tr}\left( E A E A^\dagger\right) \geq 0 \ .
\end{equation}
In particular, for $A=\gamma_\mu e^{{\rm i}
Q\hat x}$ and $A=e^{{\rm i} Q \hat x}$
we get respectively
($S^a(x)=\bar\psi(x) {\lambda^a\over 2}\psi(x)$)
\begin{eqnarray}\label{vvaa}
&& \langle V^a_\mu(Q)V^a_\mu(-Q)\rangle - \langle A^a_\mu(Q) A^a_\mu(-Q)\rangle
\geq 0 \\
&& Z(\mu)\left(\langle P^a(Q)P^a(-Q)\rangle - \langle S^a(Q) S^a(-Q)\rangle
\right)
\geq 0
\end{eqnarray}
where we used that the vector and axial current do not get renormalized
and that the scalar and pseudoscalar renormalize with the same $Z(\mu)$
factor.
The first inequality was analyzed in ref. \cite{WI} whereas the second
one (which could have been deduced from Weingarten's setting) has not been
explored in the way we shall use it.

We consider now the evaluation of the above inequalities in chiral perturbation
theory in
 the chiral limit and
in leading ${1\over N_c}$, when chiral logarithms are suppressed
\cite{GL,BRZ}. One gets \begin{eqnarray} \label{vvaach}
f_\pi^2 + 4 L_{10} Q^2 + O(Q^4) \geq 0\ ,\\ \label{vvaach2}
B_0(\mu)^2 Z(\mu) \left( {f_\pi^2\over Q^2} - 16 L_8  + O(Q^2) \right) \geq 0\
{}.
\end{eqnarray}
 It is noteworthy that all contact terms which
are related to external sources ($H_1$ and $H_2$) are
canceling in the VV-AA and PP-SS combinations. Note also that $L_{10}$
and $L_8$ appear in a scale independent combination (the same
property holds  when masses are added).
The magnitudes of $L_8$ and $L_{10}$ are such that the inequalities,
computed to second order, are indeed obeyed for $Q^2$
(Euclidean) up to $m_\rho^2$.
Therefore no rigorous signs for $L_8$ and $L_{10}$
follow from the above equations.
To get more constraining information, we would need to obtain an inequality
involving
 the momentum derivatives of the  correlators. This issue is currently under
study.

 Small variations
 of Witten's inequalities yield
\begin{equation}
 \langle P^a(Q)P^a(-Q)\rangle - \langle S^a(Q) S^a(-Q)\rangle\leq
 \langle P^a(0)P^a(0)\rangle - \langle S^a(0) S^a(0)\rangle\ ,
\end{equation}
which is trivially verified due to the presence of the pion pole (also valid
in the massive case). It is also simple to obtain the coordinate space
inequality (where no sum over $i$ is implied here)
\begin{equation}
\langle V^a_i(x)V^a_i(0)\rangle - \langle A^a_i(x) A^a_i(0)\rangle
\leq Z(\mu)\left(
 \langle P^a(x)P^a(0)\rangle - \langle S^a(x) S^a(0)\rangle\right) \ ,
\end{equation}
which is consistent with  Weingarten's inequalities.

In general, the exploitation of Witten's inequalities is  subtle.
Their information stems from unitarity constraints of the particular
$E$ operator. Physically they mingle opposite sign contributions
from many resonances. It is reasonable to expect that
they
may lead to more constraining inequalities for the physical
parameters of the chiral perturbation expansion, when corrections are
considered.

Let us finish this section with a comment on Kaplan-Manohar symmetry.
Both Eq. (\ref{mb2}) and (\ref{vvaach2}) do break the hidden symmetry of
the order $p^4$ chiral lagrangian discussed in ref. \cite{KM}.  This
reflects that the inequalities stem from QCD and, thus, tell apart
different values of, e.g., $L_8$.

\section{Some extra results}

The exploitation of QCD inequalities remains bounded to  variations
of Weingarten's and Witten's ideas. We here propose a few new
avenues for research.

Amplitudes involving  path ordered Wilson lines are easily amenable
to inequality analysis. By using a combination of Cauchy-Schwarz and
H\"older inequalities we have found that
\begin{equation}
\left|\langle\bar\psi_x \Gamma {\cal U}_{x,0} \psi_0\rangle\right|
\leq \left| \langle P^a(x)P^a(0)\rangle\right|^{1\over 2}\ ,
\end{equation}
where ${\cal U}_{x,0}= P e^{\int_0^x dz^\mu A_\mu}$ and
$\Gamma$ stands for any combination of Dirac gamma matrices.
 At long distances
hadronization will make the (gauge invariant) l.h.s. decay faster
than any meson. Nevertheless, we find it interesting that a sort of
gauge invariant constituent
mass associated to the quark line must be heavier than half of the
pion mass.

A second example of new inequality can be obtained by arguing that
the correlator of
the trace of the fermionic stress tensor is bounded by the trace of
the total stress tensor. This leads to
\begin{equation}
\langle \theta^{fermions}(x) \theta^{fermions}(0)\rangle
\leq \langle \theta^{total}(x) \theta^{total}(0)
\rangle\ .
\end{equation}
If we use chiral perturbation theory we obtain
\begin{equation}
m_q B_0 \leq m_\pi^2
\ ,
\end{equation}
which is indeed obeyed at leading order.

Let us finish by stating that no violation of the exact  QCD inequalities
that we have analyzed has been found. All current  numerical values of the
chiral perturbation theory parameters fall
in the right place. More ingenuity is necessary to produce more severe
constraints.

\section{Acknowledgments}

We are grateful to D. Espriu for discussions, A. Manohar for a relevant
comment, R. Tarrach for reading the manuscript
 and, especially, to E. de Rafael for  sharing his thorough
understanding of the subject and shaping up our ideas. G. Ecker and
M. Knecht pointed us the incorrect
 treatment of scale dependencies in Weingarten's
inequalities. Further discussions with them and with many of the participants
of the Benasque Center for Physics have been very profitable for us.
J.T. acknowledges the Universit\'e d'Aix-Marseille II
for financial support as well as
the CPT (CNRS-Luminy) for the hospitality extended
to him while this work was being completed.
Financial support from CAICYT and Comisionat per Universitats i Recerca de la
Generalitat de Catalunya and MEC under contract AEN93-0695 are also
acknowledged.


\begin{thebibliography}{99}

\bibitem{GL} J. Gasser and H. Leutwyler, {\sl Ann. Phys.} {\bf 158}
(1984) 142; {\sl Nucl. Phys.} {\bf B250} (1985) 465.
\bibitem{WE} D. Weingarten, {\sl Phys. Rev. Lett.} {\bf 51} (1983) 1830.
\bibitem{VW} C. Vafa and E. Witten, {\sl Comm. Math. Phys.} {\bf
95} (1984) 257.
\bibitem{AF} M. Asorey and F. Falceto, {\sl Phys. Lett.} {\bf B206} (1988)
485; \\ {\sl Nucl. Phys.} {\bf B327} (1989)
427.
\bibitem{WI} E. Witten, {\sl Phys. Rev. Lett.} {\bf 51} (1983) 2351.
\bibitem{FJL} D.Z. Freedman,  K. Johnson and J.I.Latorre,
{\sl Nucl. Phys.} {\bf B371} (1992) 353.
\bibitem{FSS} N.H.Fuchs, H.Sazdjian and J.Stern, {\sl Phys. Lett.} {\bf B238}
(1990) 380;  {\bf B269} (1991)  183.
\bibitem{KS} M. Knecht and J. Stern,  "Generalized Chiral Perturbation
Theory", to appear in the second edition of the
Dafne Physics Handbook, L.Maiani, G. Pancheri and
N.Paver Eds., INFN, Frascati (hep-ph/9411253).
\bibitem{FNR} E.F. Floratos, S. Narison and E. de Rafael,
{\sl Nucl. Phys.} B155 (1979) 115.
\bibitem{BCS} T. Banks and A. Casher, {\sl Nucl. Phys.} {\bf B168}
(1980) 103; A.V. Smilga, {\sl hep-th} 9503049.
\bibitem{BRZ} J. Bijnens, E. de Rafael and H. Zheng,
{\sl Z.Phys.} {\bf C62} (1994) 437;\\
J. Bijnens and J. Prades, {\sl Z.Phys.} {\bf C64} (1994) 475.
\bibitem{KM} D.B. Kaplan and A.V. Manohar, {\sl Phys. Rev. Lett.}
 {\bf 56} (1986) 2004.
\end{thebibliography}
\end{document}